\documentclass[aps,prl,twocolumn]{revtex4}
\usepackage{amsmath}
\usepackage{graphicx}
\usepackage{multirow}
\usepackage{array}
\usepackage{inputenc}
\newcommand{\hoti}{Ho$_2$Ti$_2$O$_7$}
\newcommand{\dyti}{Dy$_2$Ti$_2$O$_7$}
\newcommand{\tbti}{Tb$_2$Ti$_2$O$_7$}
\newcommand{\tbsn}{Tb$_2$Sn$_2$O$_7$}
\newcommand{\mub}{$\mu_{\rm B}$}
\newcommand{\TN}{T$_{\rm N}$}
\newcommand{\tb}{Tb$^{3+}$}

\begin{document}
\author{Sylvain Petit$^{1}$, Pierre Bonville$^{2}$, Isabelle Mirebeau$^{1}$, Hannu Mutka$^3$, Julien Robert$^{1}$}
\affiliation{$^1$ CEA, Centre de Saclay, DSM/IRAMIS/ Laboratoire L\'eon Brillouin, F-91191 Gif-sur-Yvette, France}
\affiliation{$^2$ CEA, Centre de Saclay, DSM/IRAMIS/ Service de Physique de l'Etat Condens\'e, F-91191 Gif-Sur-Yvette, France}
\affiliation{$^3$ Institut Laue Langevin, 6, rue Jules Horowitz, BP 156 F-38042 Grenoble, France}
\title{Spin dynamics in the ordered spin ice Tb$_2$Sn$_2$O$_7$}
\date{\today}
\begin{abstract}

\end{abstract}
\pacs{81.05.Bx,81.30.Hd,81.30.Bx, 28.20.Cz}
\maketitle


Geometrical frustration is a central challenge in contemporary condensed matter physics, 
a crucible favourable to the emergence of novel physics. The pyrochlore magnets, with 
rare earth magnetic moments localized at the vertices of corner-sharing tetrahedra, play 
a prominent role in this field, with a rich variety of exotic ground states ranging from the 
"spin ices" \hoti\ and \dyti\ to the "spin liquid" and "ordered spin ice" ground states in \tbti\ 
and \tbsn. Inelastic neutron scattering provides valuable information for understanding the 
nature of these ground states, shedding light on the crystal electric field (CEF) level scheme 
and on the interactions between magnetic moments. We have performed such measurements 
with unprecedented neutron flux and energy resolution, in the "ordered spin ice" \tbsn. We 
argue that a new interaction, which involves the spin lattice coupling through a low 
temperature distortion of the trigonal crystal field, is necessary to account for the data.


In spin ice materials such as the pyrochlores \hoti\ and \dyti, the orientations of 
the Ising-like rare earth magnetic moments obey local constraints, the so-called ice 
rules, analogous to those which govern the hydrogen vector displacements in water ice, 
and yield the same ground state entropy \cite{ramirez00}. The ice rules do not enforce 
long range order, and many short range ordered configurations have the same ground state 
energy. Starting from these Ising configurations, a spin flip which violates the ice rules 
induces magnetic excitations akin to magnetic monopoles, propagating along Dirac strings 
\cite{castelnovo00}. The energy necessary to create a monopole is directly related to the 
Ising character, namely to the crystal electric field (CEF) scheme of the rare earth
ion considered. In the pyrochlore structure, the crystal field of Ho or Dy ions yields a 
ground doublet with strong Ising character separated from the first excited doublet by a 
large energy gap $\Delta$ of 200 and 300\,K respectively. Weakening the Ising character 
of the wave-functions and decreasing this energy gap, as it occurs in \tbti\ and \tbsn, 
where $\Delta \sim$ 10-20\,K, provides an extra degree of freedom in the spin ice 
structure by reducing the anisotropy energy. Whereas it is not clear whether monopoles 
can still be defined in such a case, this degree of freedom allows new kinds of magnetic 
orders, possibly induced by additional perturbations, to be stabilized. Presently, the spin liquid state in \tbti\ \cite{gardner99} and the ordered spin ice state in \tbsn\ \cite{mirb05}, already studied for some time, are still not fully understood \cite{molavian07,bonv11}.

\begin{figure}
\begin{center}
\includegraphics[width=9cm]{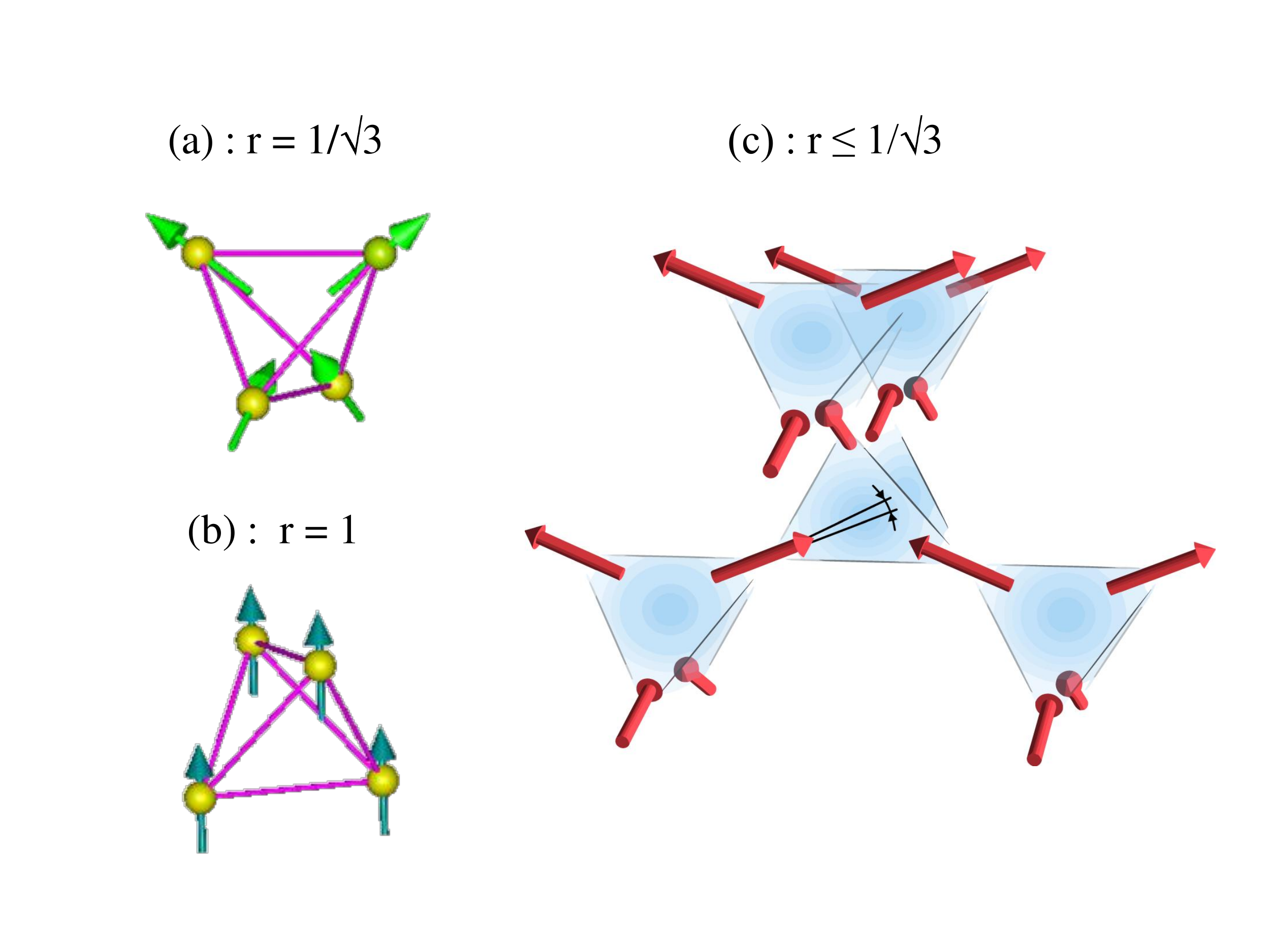}
\caption{(color on line) Different magnetic configurations on a tetrahedron : 
(a) "Two-in two out" configuration typical of the canonical "spin ice" ground state with 
magnetic moments along the local $<$111$>$ directions;
(b) Standard ordered ferromagnetic configuration; 
(c) "ordered spin ice" ground state as found in \tbsn.} 
\label{fig1}
\end{center}
\end{figure}

In this field, the ordered spin ice \tbsn\ offers a fascinating challenge. Recalling that 
an effective {\it ferromagnetic} first neighbour interaction is needed to derive the ice 
rules in the model spin ices, one can expect that, by decreasing the strength of the 
anisotropy, the short range ordered canonical spin ice, with moments along their local $<$111$>$ axes (see Fig.\ref{fig1} a), would transform into a long range ordered Heisenberg (isotropic) ferromagnet, with moments along $<$001$>$ (see Fig.\ref{fig1} b). The intermediate state for finite anisotropy would be akin to a canted non-collinear ferromagnet.
The validity of such a picture was checked by Monte-Carlo simulations \cite{champ}, where first neighbor ferromagnetic interactions in the pyrochlore lattice were combined with a uniaxial anisotropy term DS$_z^2$ of variable magnitude. In this ``soft spin ice'' model, the evolution of the magnetic structure as the anisotropy is decreased can be monitored by the variation of the moment angles with respect to their local anisotropy axes, or equivalently by the evolution of the reduced magnetisation $r = M_0/M_{\rm sat}$ along $<$001$>$. The ratio $r$ varies from 1/$\sqrt{3}=0.577$ to 1 when going from Ising to Heisenberg behaviour.  

At first sight, \tbsn\ shows most of the characteristics of this simple model. The magnetic order parameter deduced from the magnetic Bragg peak intensities \cite{mirb05} undergoes an S-shape rise as temperature decreases, with a progressive onset below 1.3\,K, an abrupt increase at \TN =0.87\,K, then a quasi-plateau below 0.6\,K. The magnetic ground state reminds that of a canted ferromagnet: the four tetrahedra of the pyrochlore cubic unit cell obey the ice rules in a coherent manner, yielding a non-zero global magnetization. For this reason, the magnetic structure found in \tbsn\ has been labelled an "ordered spin ice'' (OSI). However, a detailed inspection reveals features which do not match with the simple ``soft spin ice'' model described above \cite{champ}: i) the negative exchange contribution to the Curie-Weiss temperature \cite{mirb07} ($\theta_{\rm CW} \simeq -6$\,K), together with the first measurements of the dispersion curves of the spin excitations \cite{rule07,rule09}, strongly suggests the presence of {\it antiferromagnetic} first neighbor interactions, which is difficult to reconcile with a ferromagnetic coupling; ii) the spin structure deduced from a symmetry analysis of the neutron diffraction data and the derived canting angle of the Tb moments show that they are tilted away from the local $<$111$>$ anisotropy axis, in the direction of the $<$110$>$ axis, which leads to a spontaneous reduced magnetization $r \simeq$ 0.27 (see Fig.\ref{fig1} c) well below the range predicted by the soft spin ice model of Ref.\onlinecite{champ}. 

To study this question, we performed detailed measurements of the dynamical structure 
factor ${\cal S}(Q,\omega)$ in \tbsn\ by means of inelastic neutron scattering experiments 
(see Methods). We used the time of flight spectrometer IN5, installed on a cold neutron source at the Institut Laue Langevin (ILL), in its recent high intensity and high resolution set-up \cite{ollivier10}. Our precise data are in agreement with previous measurements \cite{rule09}; they yield the possibility to test a fine scheme of interactions, taking both the energy and intensity dependence of the magnetic scattering into account. Our measurements unveil strong similarities with the titanate \tbti\ which, despite a similar CEF scheme, remains in a spin liquid state with no magnetic order down to $T$=0 \cite{gardner99}. Based on experimental evidences, we recently proposed that a strong 4f electron - strain interaction \cite{mams}, resulting in a small "hidden" distortion developing at low temperature which breaks the local trigonal symmetry, may be a new promising route to understand this puzzling material \cite{bonv11}. The distortion induces a two singlet ground state, and we found that the spin liquid behavior arises from quantum spin fluctuations between the singlets. In the present case of \tbsn, we assume that the same physics is also relevant and examine the influence of a similar "hidden" distortion. Indeed, to account correctly for the neutron data, and especially for the strong intensity of the lowest energy excitation, we show that one needs to consider not only the exact trigonal crystal field scheme \cite{mirb07} and the interionic interaction energy term, but also the distortion from trigonal symmetry. Our model reproduces both the dynamic structure factor ${\cal S}(Q,\omega)$ and the static properties in the magnetic phase, i.e. the ordered spin ice structure with the correct canting angle and modulus of the spontaneous Tb moment. 

\begin{figure}
\begin{center}
\centerline{\includegraphics[width=8cm]{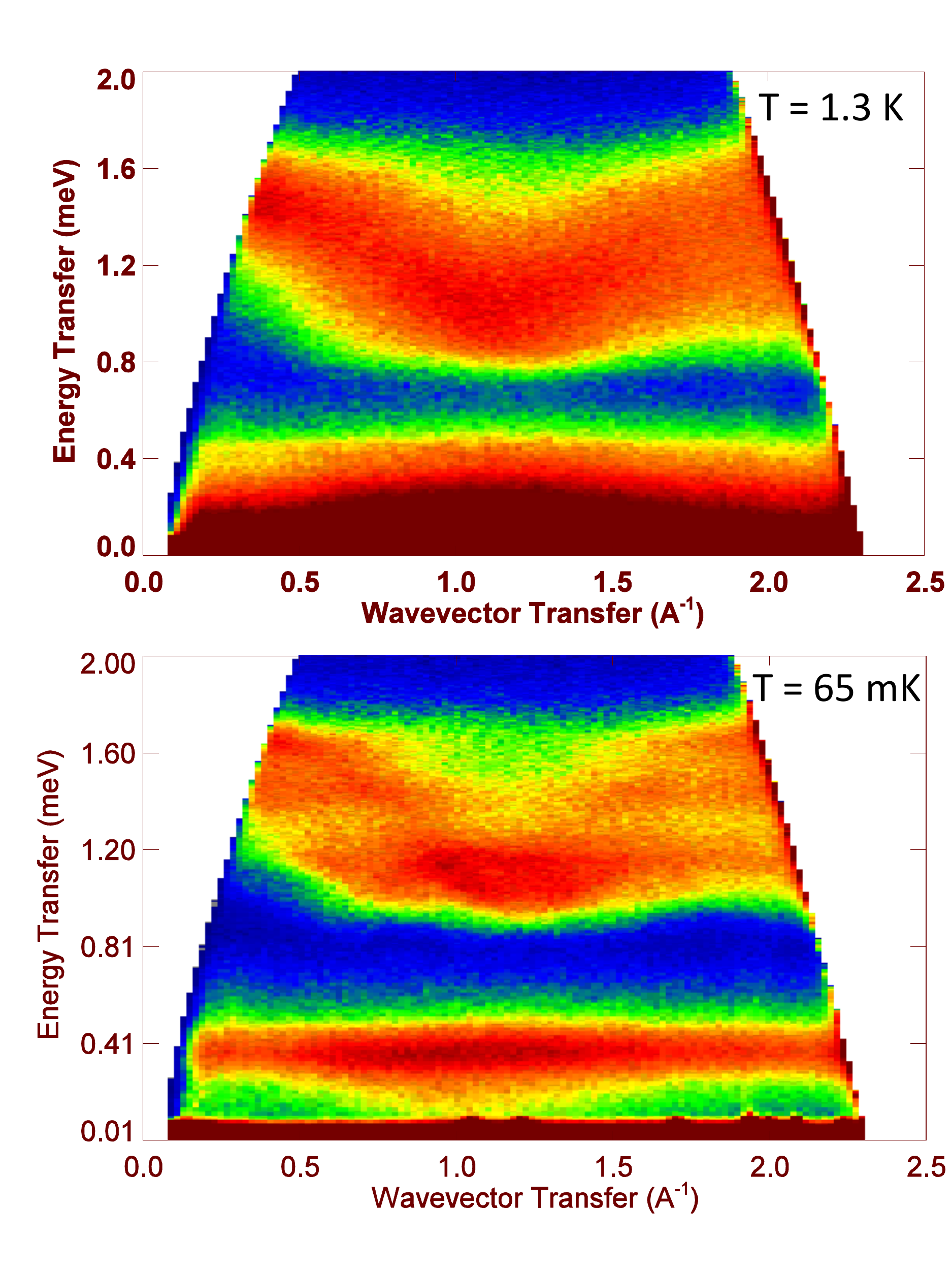}}
\caption{(color on line) : Mapping of the powder average ${\cal S}(Q,\omega)$ in \tbsn\ measured at 1.3\,K (top) and 65\,mK (bottom). These data were taken on the IN5 spectrometer installed at ILL, France, with an incident wavelength $\lambda$= 5\,\AA.}
\label{fig2}
\end{center}
\end{figure}

The upper panel of Fig.\ref{fig2} shows a mapping of the powder average ${\cal S}(Q,\omega)$ taken at T= 1.3\,K in the paramagnetic regime, just above the transition. One observes a quasielastic signal at energies less than 0.4\,meV, a gap in the 0.4-0.9\,meV region and a broad energy band in the 0.9-1.6\,meV region. This band is strongly modulated with a minimum around 1.1\,\AA$^{-1}$. All these features recall those observed in the spin liquid \tbti\ \cite{gardner01,kao,mirb07}. The modulation is due to antiferromagnetic correlations of \tb\ moments over first neighbour distances within a given tetrahedron, so that the energy of the modulation reaches a minimum nearby the liquid-like peak, whose position is related to the first neighbor distance.

\begin{figure}
\begin{center}
\centerline{\includegraphics[width=8cm]{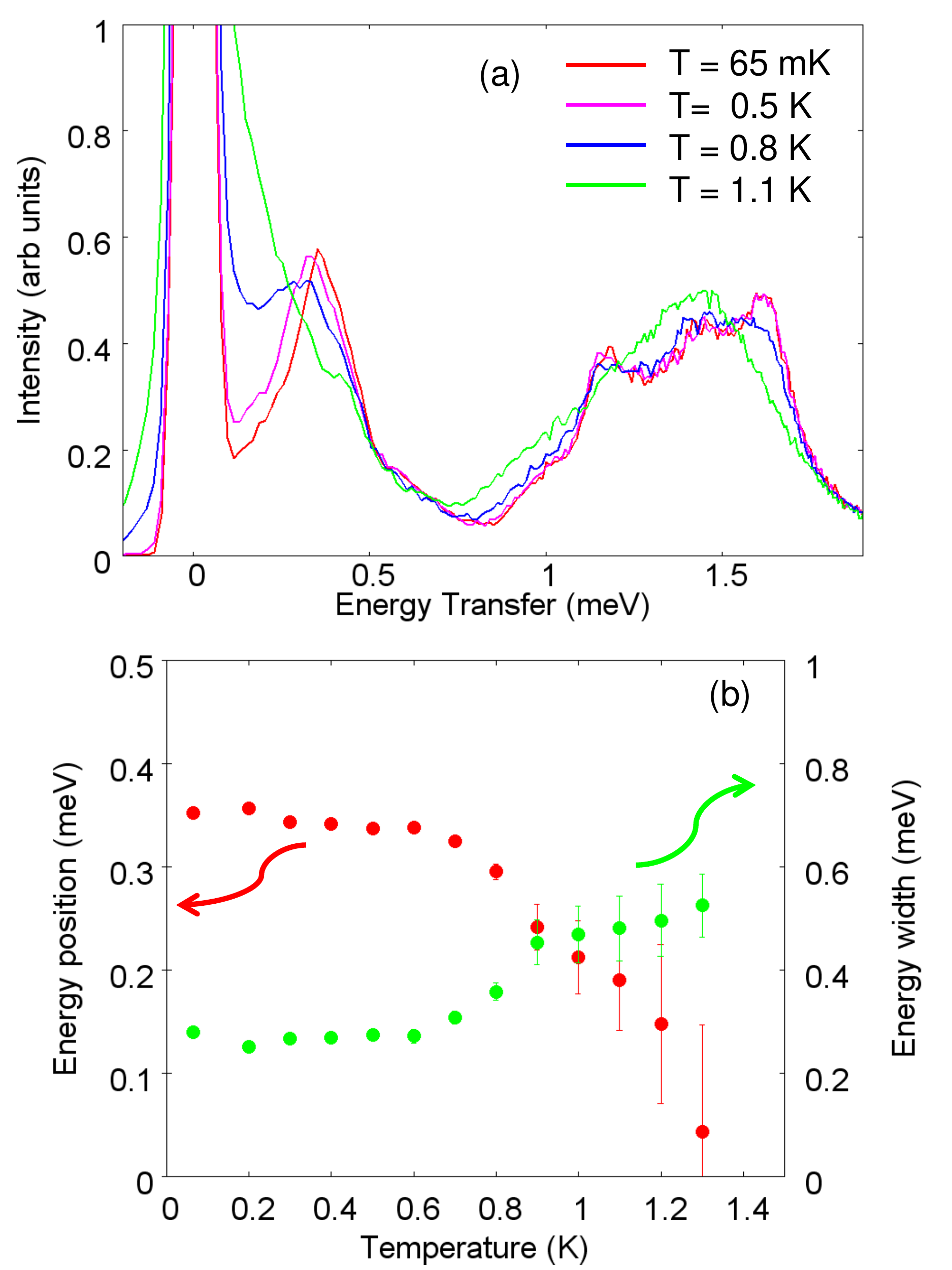}}
\caption{(color on line): (a) Energy scans for a wavevector transfer $Q$=0.5 \AA $^{-1}$ and several temperatures in the OSI phase. These cuts are obtained from the map by integrating over a $\pm$0.04 \AA $^{-1}$ wide band around $Q$.
(b) Temperature evolution of the peak position around 0.4\,meV along with its energy width, deduced from phenomenological fits performed at several temperatures.}
\label{fig3}
\end{center}
\end{figure}

At 65\,mK (lower panel of Fig.\ref{fig2}), the large quasielastic band splits into two parts. The band centered at zero energy becomes limited to a small region below 0.1 meV, where the traces of the magnetic and nuclear Bragg peaks can be detected. Well separated from it, a low energy inelastic line, which is weakly dispersive, is clearly observed around 0.4\,meV. This is further illustrated in Fig.\ref{fig3}(a), which shows several scans as a function of energy transfer and fixed $Q$ in the temperature range 65\,mK-1.1\,K. At the base temperature and up to 0.7\,K, the low energy inelastic line is well resolved. The high energy band also undergoes some changes and presents more structure than in the paramagnetic phase. Fig.\ref{fig3}(b) shows the temperature evolution of the peak position around 0.4\,meV along with its energy width. While the peak is clearly observed in the ordered phase, it becomes overdamped in the paramagnetic regime. Understanding the origin of this low energy transition appearing in the OSI phase is the main goal of the present study, and it will in turn lead to a realistic understanding of the OSI phase.

%
To analyze these experiments, one must go beyond the ``soft spin ice'' model. First, one must consider the actual CEF hamiltonian acting on a \tb\ ion ($J$=6, $g_J$=3/2): ${\cal H}_{\mbox{CEF}} = \sum_{n,m} B_n^m O_n^m$, where the $O_n^m$ are the usual Stevens operators equivalent. Second, besides the isotropic short range exchange ${\cal J}$ of a Tb ion with its $z$=6 nearest neighbours, the infinite range dipolar interaction must be included (see Methods). In the nominal trigonal symmetry, the $B_n^m$ parameters and the precise nature of the \tb\ CEF wave functions $\vert \psi_{\mu} \rangle$ and energies $E_{\mu}$ have been investigated in detail in Ref.\onlinecite{mirb07}. The \tb\ ground state is a degenerate doublet, close to:
\begin{equation}
\vert \psi_g^\pm \rangle \simeq \vert J=6; J_z=\pm 5 \rangle,
\label{doub}
\end{equation}
while the first excited state is another doublet $\vert \psi_e^\pm \rangle \simeq \vert J=6;J_z=\pm 4 \rangle$ with energy $E_e-E_g = \Delta \simeq 1.4$\,meV$\simeq 16$\,K. The dynamic structure factor ${\cal S}(Q,\omega)$ observed  in the paramagnetic phase is thus readily interpreted: the width of the quasielastic band reflects fluctuations of the Tb moments inside the ground state doublet $\vert \psi_g^\pm \rangle$, whereas the broad band at finite energy corresponds to spin excitations from the GS doublet to the first excited doublet $\vert \psi_e^\pm \rangle$ modulated by exchange. 

The main evolution of the spectra between 1.3\,K and 65\,mK, i.e. the appearance of the inelastic line at 0.4\,meV, could at first sight be explained by the trivial effect of the molecular field experienced by a Tb ion in the OSI phase. At the base temperature, when only the lowest energy level is populated, this 0.4\,meV transition would result from the exchange/dipole splitting of the GS doublet, whereas the changes in the high energy band would arise from that of the first excited doublet. The new GS eigen-functions in the OSI phase, describing the exchange split doublet, do not strongly depart from the zero field states (\ref{doub}) because of the modest magnitude of the molecular field {\bf h} (0.5-1\,T, see below). Indeed, admixtures of the excited doublet $\vert \psi_e^\pm \rangle$, which occur through the transverse component $h_\perp$ ($J_+$ and $J_-$ operators), are of order $h_\perp/\Delta$ and remain small. This point is of fundamental importance for our purpose because, for non-Kramers ions like \tb, the operator $\vec J$, which is odd with respect to time reversal, has no matrix elements between the time conjugate states of the doublet (\ref{doub}):
\begin{equation}
\langle \psi_g^+ \vert \vec J \vert \psi_g ^- \rangle  =
\langle \psi_e^+ \vert \vec J \vert \psi_e ^- \rangle  = 0, 
\end{equation}
and this also holds to a very good approximation for the exchange split states. 
Since the spectral weight of a CEF transition between states $\vert \psi_\mu\rangle$ and $\vert \psi_\nu \rangle$ observed by neutron scattering is proportional to $| \langle \psi_{\mu} \vert \vec{J} \vert \psi_{\nu} \rangle|^2$, the transition between the molecular field split states of the GS doublet is expected to have a very small intensity. Hence, the corresponding line should be extinct, contrary to experimental observation. 

\begin{figure}
\begin{center}
\centerline{
\includegraphics[width=9cm]{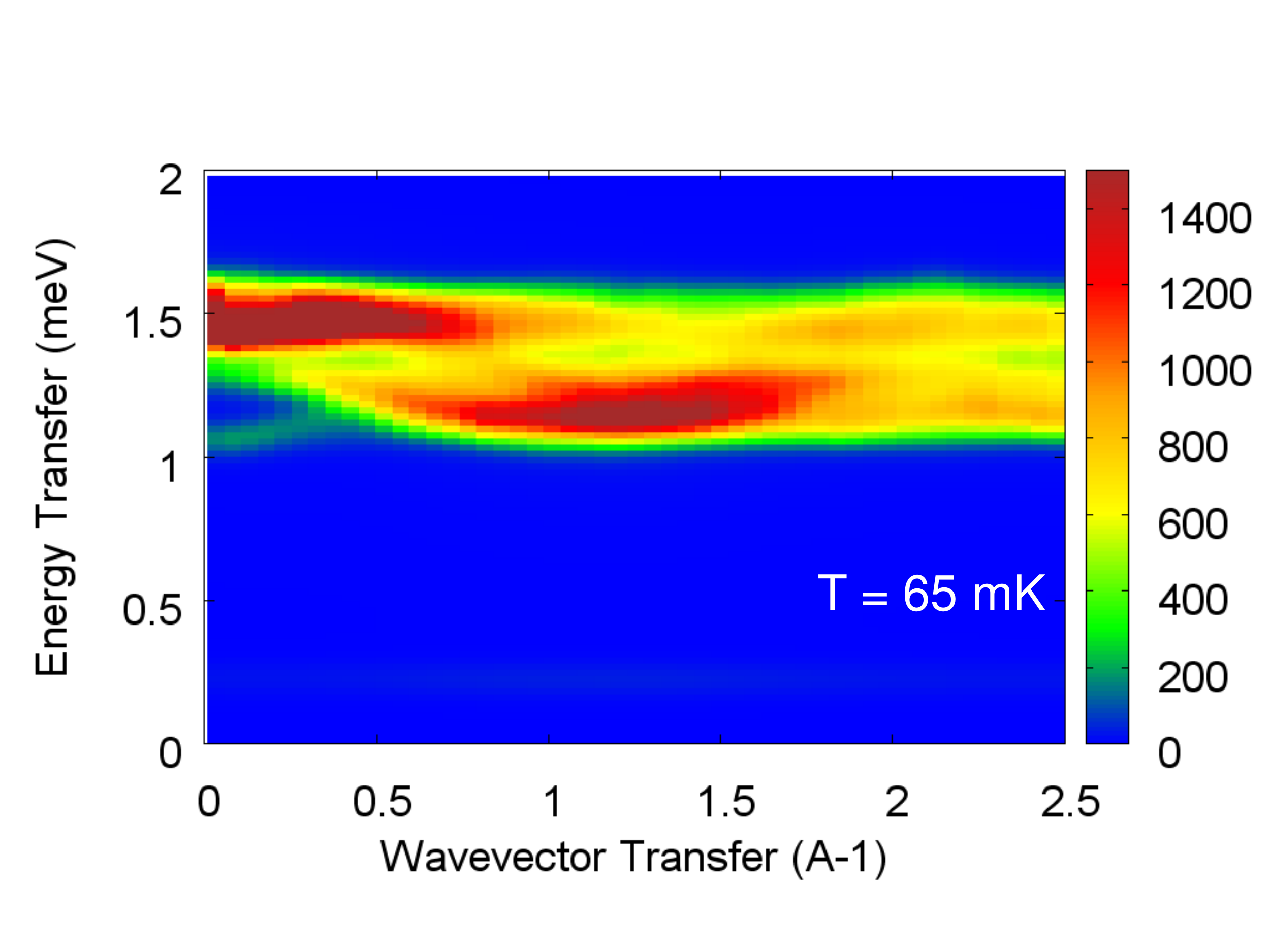}}
\caption{(color on line) Calculated powder average ${\cal S}(Q,\omega)$ at $T=65$\,mK in the RPA approximation for ${\cal J}=-0.09$\,K, using the trigonal crystal field of \tbsn.}
\label{fig4}
\end{center}
\end{figure}

To put these arguments on a more quantitative ground, we numerically diagonalise the total hamiltonian in the mean field approximation and compute ${\cal S}(Q,\omega)$ in the standard RPA approximation (see Methods), giving access to the dispersion of the collective spin excitations. In \tbsn, an antiferromagnetic exchange integral $\cal J=-$0.09\,K can be inferred from the analysis of the Curie-Weiss law (see Methods). Fig.\ref{fig4} shows that, at 65\,mK, ${\cal S}(Q,\omega)$ consists in a set of crystal field transitions around 1.5\,meV slightly modulated by the exchange/dipole coupling between magnetic moments. Interestingly, despite the ferromagnetic-like nature of the ground state, the minimum of the calculated dispersion curve does not occur at the ferromagnetic zone center ($Q$=0), but at its boundary, i.e. it is in line with the antiferromagnetic exchange integral. However, as expected, a comparison between Fig.\ref{fig4} and the lower panel of Fig.\ref{fig2} shows that the model is not able to capture the intensity of the low energy transition observed at 0.4\,meV. 

Explaining the data thus demands the introduction of a new interaction term that would imply a major change in the CEF wave functions. At this step, it is worth emphasizing that, for a non-Kramers ion, a degenerate doublet is very unstable against any crystal field perturbation. Several arguments suggest that a tetragonal distortion is at play in the sibling material \tbti: there exists clues to the presence of a $T \simeq 0$ Jahn-Teller transition \cite{mams} and of cubic-to-tetragonal structural fluctuations below 20\,K \cite{ruff07}. The distortion is likely to be at the origin of the spin liquid state of this material \cite{bonv11}. We assume here that such a distortion is also present in \tbsn\ and we examine the consequences of this hypothesis on the physical properties. 
With its principal axis along the cubic [001] axis, the distortion is described by the following crystal field operator equivalent, in the local frame with the $<$111$>$ axis as z-axis \cite{rulepb09}:
\begin{equation}
{\cal H}_Q = \frac{D_Q}{3}\ [2J_x^2+J_z^2+\sqrt{2}\ (J_xJ_z+J_zJ_x)].
\label{dq}
\end{equation}
Let's examine first the changes introduced by the distortion on the ground state in the paramagnetic phase. As in the case of \tbti, the principal effect is a degeneracy lifting of the ground doublet (and of the first excited doublet) and a deep modification of its wave-functions. Whatever the nature of the distortion, it can easily be shown that the new eigenstates become entangled singlet states of symmetric and antisymmetric type which, for the ground state, are close to:
\begin{equation}
\vert \psi_{s,a} \rangle \simeq \frac{1}{\sqrt{2}}\ [\vert J=6;J_z=5 \rangle \pm \vert J=6;J_z=-5 \rangle].
\label{ent}
\end{equation}
These states are such that $\langle \psi_s \vert \vec J \vert \psi_s \rangle = \langle \psi_a \vert \vec J \vert \psi_a \rangle$=0, but, by contrast with the non-perturbed states (\ref{doub}), the cross matrix element $\langle \psi_a \vert J_z \vert \psi_s \rangle$ is non-vanishing and large. For distortion amplitudes $D_Q$ similar to those derived in \tbti\ (0.25\,K \cite{bonv11}), the energy separation between the singlets amounts to a few 0.1\,meV (a few K). Therefore, we expect a new low energy transition in the CEF scheme from $\vert \psi_s \rangle$ to $\vert \psi_a \rangle $ with a large spectral weight. 

\begin{figure}
\begin{center}
\centerline{
\includegraphics[width=9cm]{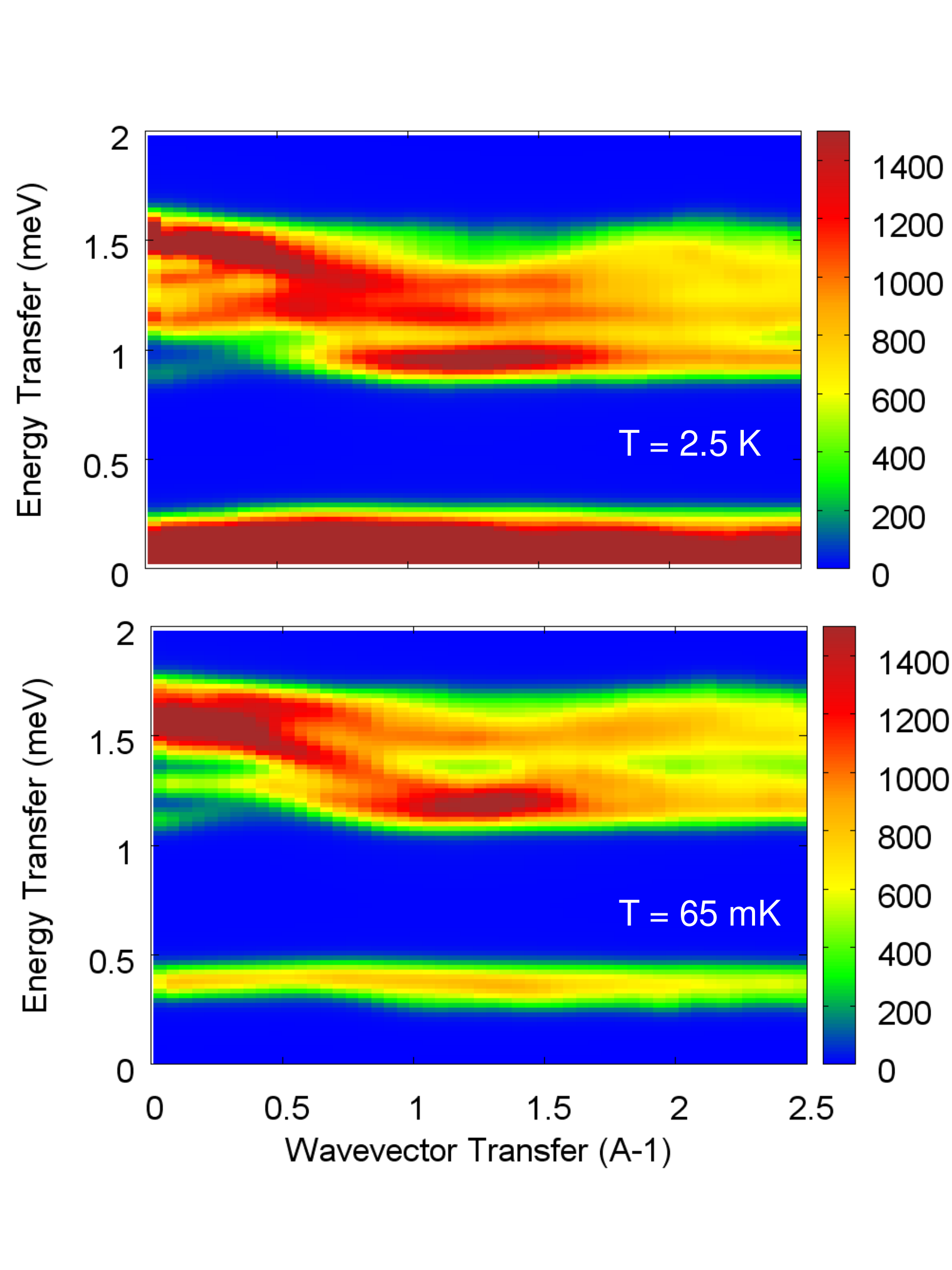}}
\caption{(color on line) Calculated powder average ${\cal S}(Q,\omega)$ with a tetragonal distortion of the trigonal CEF ($D_Q=0.2$\,K), for ${\cal J}=-0.09$\,K, in the paramagnetic phase ($T=2.5$\,K, upper panel) and in the ordered spin ice phase ($T=65$\,mK, lower panel).}
\label{fig5}
\end{center}
\end{figure}

\begin{figure}
\begin{center}
\includegraphics[width=9cm]{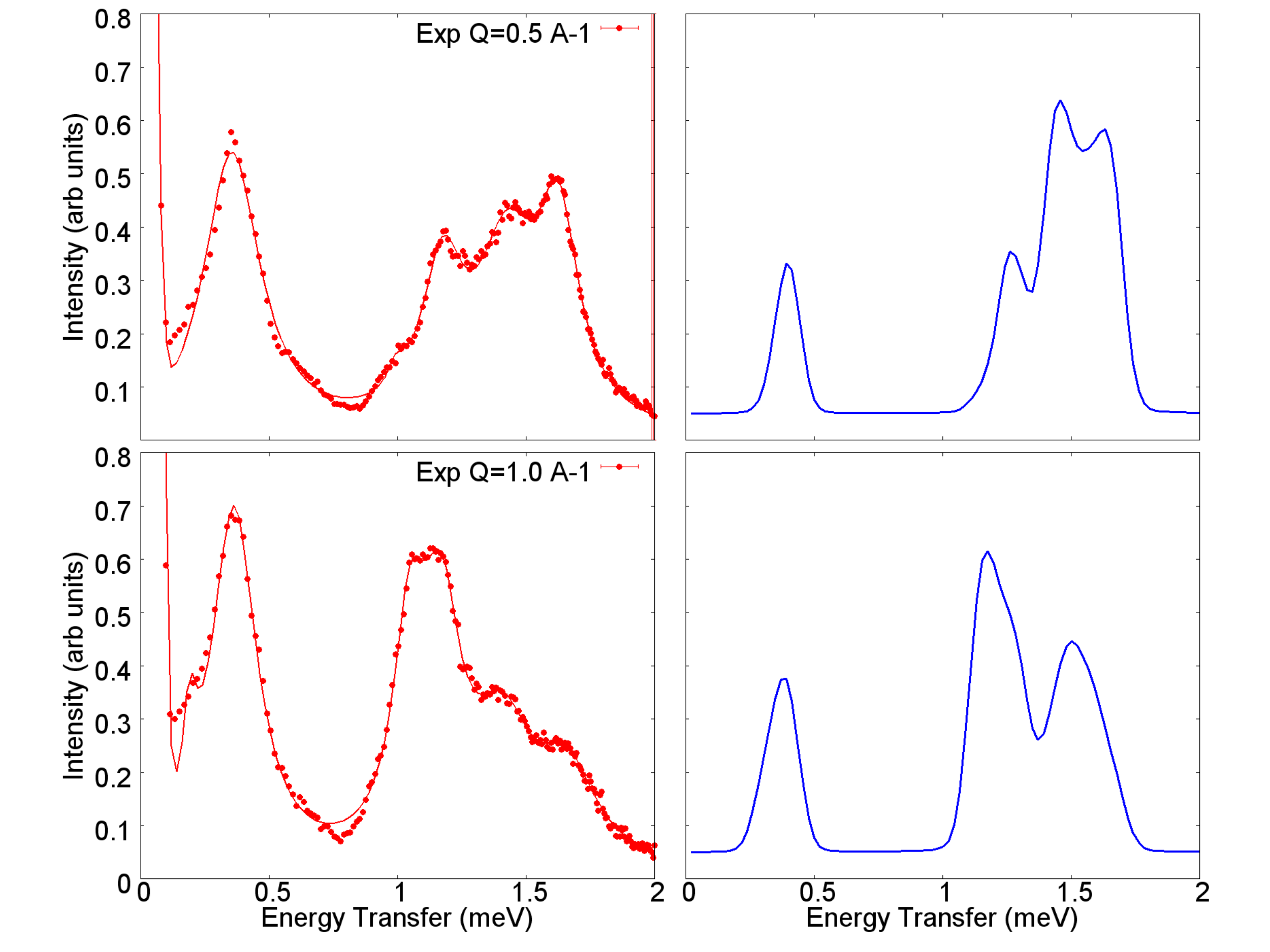}
\caption{(color on line) Experimental (left) and calculated (right) inelastic spectra at 65\,mK, for $Q=0.5$ and $1.5$\AA$^{-1}$, in \tbsn. The simulations were performed with ${\cal J}=-0.09$\,K and $D_Q$=0.2\,K. A constant background was added to the 
calculated profiles, and their overall intensity was arbitrarily scaled to reproduce the intensity of the upper band at $Q=1$\AA$^{-1}$.} \label{fig6}
\end{center}
\end{figure}

In the OSI phase, for the same reason as mentioned above, the eigen-functions do not depart strongly from the singlets (\ref{ent}), resulting in a sizeable intensity low energy transition, as observed experimentally. The intensity map calculated in the presence of the distortion for $\cal J=-$0.09\,K is represented in Fig.\ref{fig5}, in the paramagnetic phase (upper panel) and in the ordered phase (lower panel), to be compared with Fig.\ref{fig2}. Fig.\ref{fig6} shows 1D energy cuts (experimental and calculated) for two Q values at 65\,mK. As expected, at the base temperature, a sizeable intensity for the low energy line is recovered, in agreement with experiment, and its position (0.4\,meV) is correctly reproduced with a distortion strength $D_Q=0.2$\,K. Its intensity, however, is lower than observed. Preliminary calculations show that the agreement with experiment can be improved by introducing anisotropic exchange, but this will not be discussed here. The $Q$-dependence of the structure in the upper broad band is also rather well captured by the calculation, at a semi-quantitative level reproducing the inversion of the energy asymmetry between $Q$=0.5 and $Q$=1.0\AA$^{-1}$. In the paramagnetic phase, for $D_Q=0.2$\,K, the splitting between the two singlets should be $\delta \simeq 0.17$\,meV. The corresponding inelastic line is thus expected to merge into the quasi-elastic peak, which extends up to 0.4\,meV at 1.3\,K, and this explains why it cannot be clearly observed in the experiment. The above $\delta$ value is compatible with the thermal evolution of the energy of the inelastic line shown in the lower panel of Fig.\ref{fig3}, which reaches 0.17\,meV at 1.2\,K and becomes overdamped above this temperature.

It seems therefore clear that the introduction of a tetragonal distortion greatly improves the agreement between the calculated dynamic structure factor and the inelastic neutron data. We turn eventually to the static properties of the OSI phase in \tbsn\ by computing the magnetic phase diagram (${\cal J},T_{\rm C}$) using our 4-site self-consistent model (see Methods). Without distortion, as expected, large negative (AF) $\cal J$ values (${\cal J} \le {\cal J}_c=-0.22$\,K) stabilize a non-collinear N\'eel ground state, the so-called "all in - all out" antiferromagnetic structure. Above this critical value, the ground state is the ordered spin ice phase (OSI) with the same characteristics as observed by neutron diffraction in \tbsn\ \cite{mirb05}. The existence of this ferromagnetic-like phase with an AF exchange integral is fully consistent with the now well established idea that, in the presence of dipolar interactions, a spin ice ground state can be stabilized for moderate AF exchange \cite{hertog2000}. This result is also consistent with the ``soft dipolar spin ice'' model \cite{clarthy10} recently proposed to account for the ground state of \tbsn. Our mean field model reproduces a peculiarity of the OSI phase, already mentioned in the introductive section, that puts it outside the reach of the "soft spin ice" model of Ref.\onlinecite{champ}: the magnetic moment is found to tilt away from the $<$111$>$ axis, in the direction of the $<$110$>$ axis, or equivalently the reduced magnetisation $r$ is lower than 1/$\sqrt{3}$.
\begin{figure}
\begin{center}
\includegraphics[width=8cm]{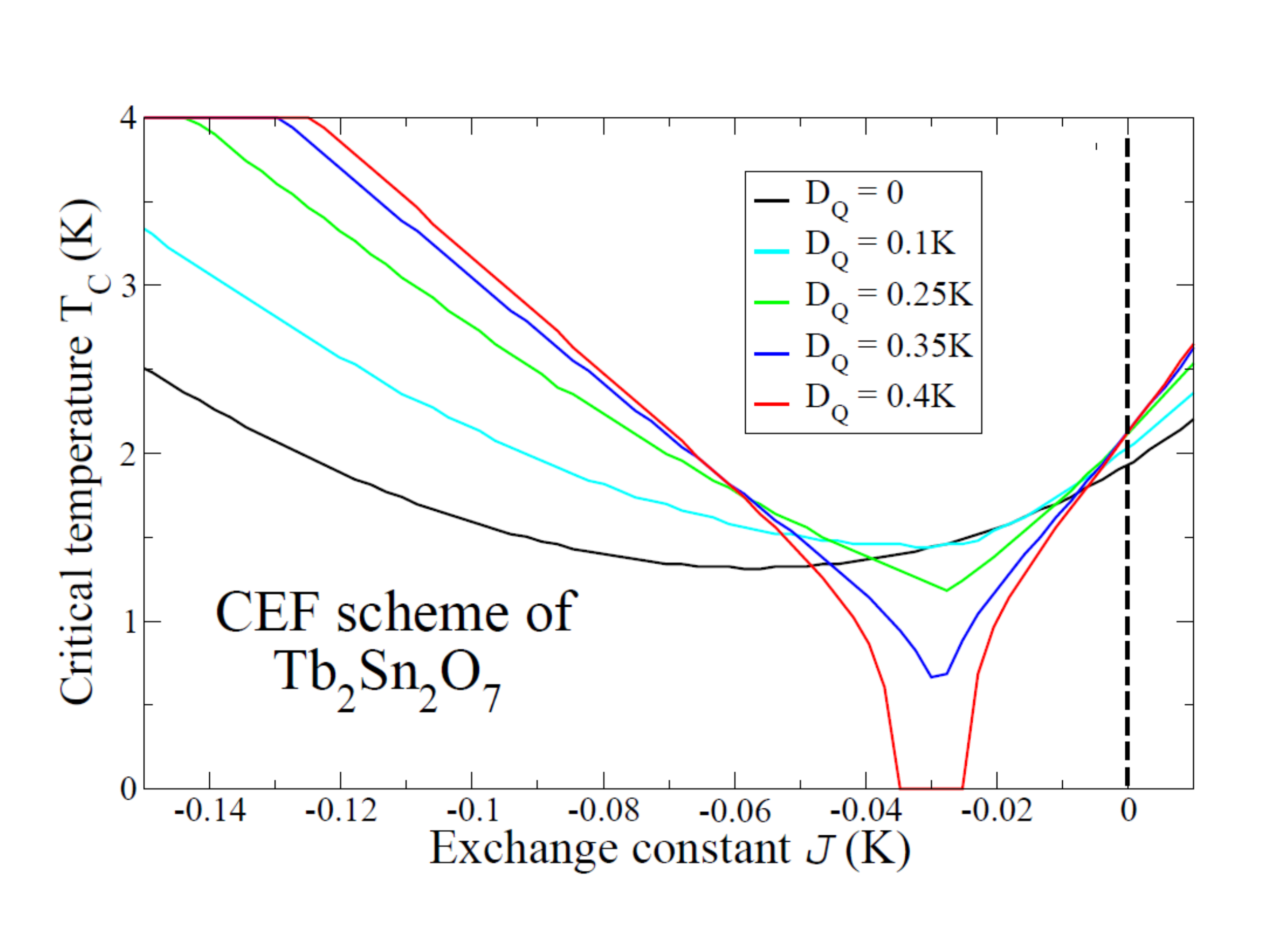}
\caption{(color on line) Ordered spin ice part of the (${\cal J},T_{\rm C}$) phase diagram computed using the crystal field interaction in \tbsn\ in the presence of the tetragonal distortion (\ref{dq}), for different values of the $D_Q$ parameter.} 
\label{diagphas}
\end{center}
\end{figure}
Figure \ref{diagphas} focuses on the OSI part of the phase diagram, computed for different amplitudes $D_Q$ of the distortion. For $D_Q \le 0.25$\,K, the critical temperature $T_{\rm C}$ shows a regular variation with $\cal J$. For the parameter values characteristic of \tbsn: ${\cal J}=-0.09$\,K and $D_Q \simeq 0.2$\,K, we find a transition temperature $T_{\rm C} \simeq$2.35\,K, higher than the experimental value 0.9\,K \cite{mirb05}. We think this difference is due to a possible first order character of the transition at 0.9\,K, suggested by the abrupt S-shape variation of the magnetic Bragg intensity near $T_{\rm C}$. The saturated Tb moment is 6.15\,\mub, close to the experimental value 5.9(1)\,\mub, and the angle between the moment and the $<$111$>$ axis is 15$^\circ$, also close to the experimental value 13$^\circ$. The total molecular field (exchange + dipolar) is $h=0.78$\,T and it makes an angle of 55$^\circ$ with the local $<$111$>$ axis.

For higher distortion amplitudes, a dip appears in the $T_{\rm C}(\cal J)$ curve for low $\vert \cal J\vert$ values, and above a distortion amplitude $D_Q \simeq 0.4$\,K, one observes a strong depletion of the ordered spin ice critical temperature, which goes to zero in a narrow range around ${\cal J} \simeq -0.03$\,K. This phenomenon is reminiscent of the problem of the singlet ground state in non-Kramers rare earth systems \cite{bleaney}, where no magnetic order occurs down to $T=0$ if the exchange integral is below a critical value proportional to the gap separating the singlet from the excited states. In a recent work \cite{bonv11}, we argue that this case is realized in the spin liquid \tbti. Likewise, a sizeable intensity low energy transition appears in the neutron scattering spectra at low temperature, both in zero field and in the presence of an external field, which cannot be indexed in the trigonal CEF/Zeeman level scheme \cite{rulepb09,bonv11}. Our picture points to the essential role of the precise CEF scheme of \tbsn\ and \tbti, as well as to the relevance of a new interaction, namely the 4f electron - strain coupling, to understand the properties of these compounds. In this sense, our model is very different from the many body approach described in \cite{molavian07}, which emphasizes the importance of virtual crystal field excitations. 

In summary, we have measured low energy spin excitation spectra in the  "ordered spin ice'' pyrochlore \tbsn\ with a very high resolution. In the frame of a mean field calculation, we have shown that the ferromagnetic-like ordered spin ice structure is the ground state in this material for moderate {\it antiferromagnetic} exchange and taking into account dipolar interactions between Tb moments. We also find that the introduction of a small lattice distortion from trigonal symmetry is necessary to explain most features of the low temperature dynamical structure factor, in particular the existence of a low energy transition in the ordered spin ice phase. Indeed, due to a specific property of non-Kramers doublets, the neutron scattering intensity between trigonal symmetry states split by exchange is vanishingly small. The perturbation brought up by the distortion results in  "entangled'' tunnel-like wave-functions for the two lowest levels, yielding a sizeable intensity inelastic line, as observed. Our model reproduces the details of the magnetic structure in \tbsn, and our results strongly suggest that the 4f electron - strain coupling is a key feature to analyze the quantum spin ice states in terbium pyrochlores. The physics underlying the "opposite'' behaviours of the sibling materials \tbti\ and \tbsn, i.e. respectively the spin liquid and the ordered spin ice phase as ground state, appears to be described by the same model, but with slightly different parameters. 

\section{Aknowledgments}

We thank Jacques Ollivier for his technical support during the neutron experiment on IN5, Anne Forget for preparing the sample and Andrey Sazonov for support with the picture design. 

\section{Methods}
\subsection{Experimental details}

A powder sample of 2.2\,g, the same as used in Ref.\onlinecite{mirb05}, was inserted in a
rectangular aluminium sample holder to minimize the absorption corrections, then placed
into a dilution insert inside a helium cryostat. The temperature was varied from
65\,mK to 1.6\,K. For the measurements on the IN5 spectrometer, we used three incident 
wavelengths of neutron beam: $\lambda$ = 5, 8 and 10\,\AA, yielding respectively an 
energy resolution of 90, 20 and 15\,$\mu$eV, at full width half maximum (FWHM) at the 
elastic peak. Spectra were corrected from the background of the empty sample holder and 
from neutron absorption. The detector efficiency was taken into account by measuring a 
vanadium plate of the same dimensions as the sample. The energy resolution was determined 
with a vanadium foil inserted in a cylindrical sample holder.

\subsection{Mean field calculation}

The total mean field Hamiltonian for a \tb\ ion at a given site $i$ reads:
\begin{equation}
{\cal H} = {\cal H}_{\mbox{CEF}} + {\cal H}_Q - \vec{J}_i
(\sum_{j=1}^z {\cal \tilde J}_{i,j}+\sum_j^\infty{\cal \tilde J}\mbox{dip}_{i,j})
\langle \vec{J}_j \rangle
\end{equation}
where $\tilde {\cal J}_{i,j}$ = ${\cal J}$ denotes the isotropic short range exchange of a Tb moment with its $z$=6 nearest neighbours, and ${\cal \tilde J}\mbox{dip}$ corresponds to the infinite range dipolar interaction calculated using Ewald summation \cite{ewald}. In defining the neighborhood of a given moment, it is worth emphasizing that we assume a $\vec k$=0 magnetic structure in the $Fd\bar 3m$ space group with face centered cubic (fcc) symmetry. In other words, the 4 Tb moments of a given tetrahedron may be different, 
but the spin configurations on tetrahedra connected by fcc lattice translations are the same. We consider the 4 moments at the vertices of a tetrahedron, diagonalize $\cal H$ in the local frame for each site to determine the energies $E_\mu$ and wave functions $\vert \psi_\mu \rangle$, and perform a self-consistent calculation to obtain the \tb\ magnetic moments $-g_J \mu_B \langle \vec{J}_i \rangle$ at a given temperature. The only parameters are the exchange integral $\cal J$, the lattice parameter $a$ and the distortion magnitude $D_Q$. The AF exchange integral in \tbsn\ is ${\cal J} = -0.09$\,K, a value derived from the molecular field constant $\lambda$ obtained in Ref.\onlinecite{mirb07}, using the relationship:
\begin{equation}
{\cal J} = \frac{1}{6} \lambda \frac{g_J^2 \mu_B^2}{k_{\rm B}}.
\end{equation}

\subsection{RPA calculations}

In the RPA approximation, ${\cal S}(Q,\omega)$ is given by the spin-spin susceptibility :
\begin{eqnarray*}
{\cal S}(Q,\omega) & = & \sum_{m,n}~e^{i\vec Q(\vec r_m-\vec r_n)} \\
& & \sum_{\alpha, \beta=x,y,z}
\left(\delta_{\alpha \beta}-\frac{Q^\alpha Q^\beta}{Q^2}\right)~\chi^{\alpha \beta}_{m,n}(\omega),
\end{eqnarray*}
where $\vec r_m$ denote the position of spin at site $m$. The $3 \times 3$ tensor $\tilde \chi_{i,j}$ is solution of the equation:
\begin{eqnarray*}
\chi^{\alpha \beta}_{i,j} & = & \xi^{\alpha \beta}_{i} \delta_{i,j} - \sum_{\alpha',\beta'=x,y,z}\xi^{\alpha \alpha'}_{i}~\times \\
& & \sum_{\ell}~
\left({\cal J}^{\alpha' \beta'}_{i,\ell}+{\cal J}\mbox{dip}^{\alpha' \beta'}_{i,\ell}\right)~\chi^{\beta' \beta}_{\ell,j}.
\end{eqnarray*}
The $\tilde \xi_i$ tensor is the $3 \times 3$ bare susceptibility, whose elements $\xi^{\alpha \beta}_{i}$ are given by :
\begin{equation}
\xi^{\alpha \beta}_{i}(\omega) = \sum_{\mu,\nu}~
\left(n_{\mu}-n_{\nu} \right)~
\frac{W^{\alpha}_{i \mu \nu}~W^{\beta}_{i \nu \mu}}
{\omega+i0^ + E_{\nu} - E_{\mu}}.
\end{equation}
The calculation relies on the eigen-energies $E_{\mu}$ and eigenvectors $\vert \psi_{i \mu} \rangle$ ($i$ is a site index) determined in the $\vec k$=0 mean field approximation. $n_{\mu}$ denote the occupation of level $\mu$, $n_{\mu}= (\exp{-E_{\mu}/k_{\rm B}T})/Z$. The $\vec{W}_{i \mu \nu}$ are vectors $(W^x_{i \mu \nu}, W^y_{i \mu \nu}, W^z_{i \mu \nu})$ corresponding to the matrix elements of the total angular momentum $\vec{J}_{i}= \sum_{\mu,\nu} \vec{W}_{i \mu \nu} \vert \psi_{i \mu} \rangle \langle \psi_{i \nu} \vert$ in the eigen-basis.


\end{document}